\pdfoutput=1
\documentclass{article}

\usepackage{PRIMEarxiv}

\usepackage[utf8]{inputenc} 
\usepackage[T1]{fontenc}    
\usepackage{hyperref}       
\usepackage{url}            
\usepackage{booktabs}       
\usepackage{amsfonts}       
\usepackage{nicefrac}       
\usepackage{microtype}      
\usepackage{fancyhdr}       
\usepackage{graphicx}       
\usepackage{amsmath}        
\usepackage{float}          

\title{What FLARE's Published Numbers Actually Measure}

\author{
  Dung Cao \\
  Optivex Research \\
  \texttt{ddcao09@gmail.com} \\
  \And
  Palaash Gang \\
  Indus International School, Pune \\
  \texttt{palaashgang@gmail.com} \\
  \And
  Arjun Veluri \\
  Indus International School, Pune \\
  \texttt{arjunsveluri@gmail.com} \\
}

\begin{document}
\maketitle

\begin{abstract}
We audit the scoring pipeline of FLARE, the financial task suite behind PIXIU and FinBen, and find that in a decisive fraction of cells the score is a property of the answer-extraction rule, not the model. The clearest evidence is already published: of the 34 released \textsc{Headlines} scores, nine equal to four decimals what the benchmark's own code returns when no output is byte-equal to \texttt{Yes}, and only four clear the per-stratum majority floor. FLARE credits the first entry of the dataset card's \texttt{choices} list found anywhere in a generation; the \textsc{lm-evaluation-harness} it forks takes the first match by position; on identical generations the two differ by up to 0.4216 accuracy, and no leaderboard states which rule produced its numbers. An unspecified rule leaves the score partially identified, and because the coarsening is multiplicity rather than missingness the interval's width is counted rather than bounded. Across a 204-cell factorial and a 46-cell diagnostic, 10 cells are certified invariant to the rule, 4 only because almost nothing parses, so certification is paired with coverage. On Financial PhraseBank four of seven models fall below the 0.5948 majority floor under one shipped rule and clear it under the other; on three of four substring-scored tasks no open-weight model we test is distinguishable from a degenerate predictor. Reading a FLARE score needs three quantities it does not report -- ambiguity rate, coverage, floor -- and \textsc{floorbench} computes all three from generations already produced.
\end{abstract}

\par\addvspace\medskipamount{\rightskip=0pt plus1cm\noindent{\bfseries\emph{CCS Concepts}}\enspace\ignorespaces
\textbullet~\textbf{Computing methodologies} $\rightarrow$ \textbf{Machine learning}; \textbf{Natural language processing}; \textbullet~\textbf{Information systems} $\rightarrow$ \emph{Information systems applications}.\par}

\keywords{financial benchmarks, LLM evaluation, reproducibility, answer extraction, trivial baselines}

\section{Introduction}\label{sec:intro}
\begin{figure}[t]
  \centering
  \includegraphics{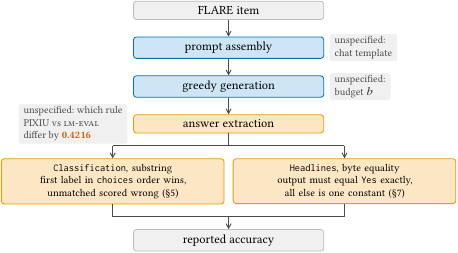}
  \caption{The audited pipeline, with the parameters no leaderboard reports tagged beside the stage each one governs. Extraction forks into two shipped rules that fail differently, and both feed the same reported number.}
  \label{fig:pipeline}
\end{figure}

This paper audits the scoring pipeline of one financial benchmark. FLARE, the task suite behind PIXIU and FinBen \cite{finben,pixiu}, ships its classification tasks as instruction prompts and scores them by extracting a label from generated text. That extraction rule is a free parameter. FLARE's \texttt{Classification} path returns the first entry of the dataset card's \texttt{choices} list occurring anywhere in the lowercased generation; the \textsc{lm-evaluation-harness} it forks \cite{biderman} returns the first match by position. On identical generations the two differ by up to \textbf{0.4216} accuracy, and no leaderboard reports which one produced the number. Figure~\ref{fig:pipeline} marks that fork and the two other parameters the benchmark leaves unspecified, the prompt arm and the generation budget.

We ran the pipeline as a 204-cell factorial (6 open-weight models, 5 FLARE tasks, 2 prompt arms, 4 budgets, 364,260 generations, both extraction paths). Three things follow. Extraction ambiguity is large enough to reverse conclusions: on Financial PhraseBank four of seven models fall below the majority-class floor under one shipped rule and clear it under the other, on the same generations (the diagnostic's seven-model roster; \S\ref{sec:design}). A floor-level score does not say why it sits there: on German credit a model that answers and loses, one whose score the parser assembles from a restated question (\S\ref{sec:echo}), and one that never parses sit at one floor, ranked backwards (\S\ref{sec:floor}). And the signature is already in the published record: of the 34 released \textsc{Headlines} scores, nine are published at four decimals equal to the value the benchmark's own code returns when no output is ever byte-equal to \texttt{Yes} (\S\ref{sec:exact}).

To say how exposed a given cell is we use an inherited bound rather than a new theorem. An unspecified extraction rule leaves the reported score \emph{partially identified}, and the interval it is pinned to is the standard worst-case identification bound \cite{manski1989,manski2003}, which Rao and Callison-Burch have already brought to LLM evaluation \cite{rao}. What this pipeline contributes is a coarsening of a different kind, multiplicity rather than missingness, whose candidate set the harness already computes, so the statistic is counted rather than bounded (\S\ref{sec:rule}). We report it beside coverage throughout, because a cell that parses nothing is certified for the wrong reason.

The audit belongs at a finance venue because what it puts at risk is a specific body of financial numbers: the FLARE scores now cited and compared against were produced by this pipeline, and newer tooling changes future runs without reinterpreting figures already in print. The findings do not transfer to another suite; the method and \textsc{floorbench} do. Nothing here depends on a dataset being wrong: every finding is compatible with every label being correct.

The exposure is not a quirk of the audited commit: answers on 46 of the suite's 80 task classes resolve by substring (\S\ref{sec:background}), and it survives the chat template -- \textbf{13} of the 18 chat-arm cells stay exposed -- and the extraction filter the current release ships (\S\ref{sec:rule}). Neither repair we test serves both extraction paths (\S\ref{sec:recover}), so the remedy is reporting rather than repair: \textsc{floorbench} computes coverage, ambiguity and the per-task floor from generations a run has already produced (\S\ref{sec:spec}).

\begin{table}[t]
  \centering
  \footnotesize
  \begin{tabular}{@{}lr@{}}
    \toprule
    task classes in \texttt{flare.py} & 80\\
    \midrule
    inherit \texttt{Classification} (39 of them directly) & 48\\
    \quad reach \texttt{process\_results} unmodified & 37\\
    \quad \texttt{StockMovement} override: synonym table $+$ \texttt{DEFAULT} & 9\\
    \quad \texttt{Headlines}/\texttt{ZHHeadlines}: byte-equality replacement & 2\\
    resolve by substring (37 $+$ 9) & 46\\
    generative and sequence-labelling, own scorers & 32\\
    \bottomrule
  \end{tabular}
  \caption{Census of scoring paths at the audited commit.}
  \label{tab:census}
\end{table}

\section{Background and Related Work}\label{sec:background}
\paragraph{The audited pipeline.} FLARE tasks ship as instruction prompts with a fixed answer vocabulary; the harness generates greedily to a budget, then extracts a label by substring (\texttt{Classification}) or by byte equality (\texttt{Headlines}). At the audited commit (\texttt{b92bc75}, unchanged since 2025-03-04) the 80 task classes in \texttt{flare.py} partition as Table~\ref{tab:census} shows: 46 of the 48 \texttt{Classification} inheritors resolve by substring, only 37 of them by the audited code. Three of our four sweep tasks sit in the \texttt{StockMovement} override (\texttt{German} and \texttt{Australian} with \texttt{DEFAULT} $=$ \texttt{good}, \texttt{StockMovementCIKM} with \texttt{fall}); only \texttt{FPB} reaches \texttt{Classification.process\_results}, and \S\ref{sec:floor} shows what the override does to a score. Prompt assembly and generation budget are call-site parameters with no task-level default.

\paragraph{Trivial floors and their treatment in FinBen.} Comparing against degenerate baselines is an old lesson \cite{gururangan}, and the frequent failure of LLMs to beat a majority-class baseline is in print: on uncontaminated classification tasks they rarely improve significantly on simple majority baselines \cite{li2024}, and a majority-floor re-evaluation of tabular models reaches the same verdict \cite{gorla}. FinBen states the adjacent observation itself (weak models ``tend to classify all cases into a single class'') but attributes it to ``low instruction-following abilities'' and task difficulty \cite{finben}. We attribute it to the scoring path, show it afflicts strong and weak models alike, and supply the floor the benchmark omits; ``majority,'' ``base rate,'' ``naive'' and ``constant'' occur in neither it nor PIXIU \cite{pixiu} (arXiv:2402.12659v2 / arXiv:2306.05443v1; \texttt{paper/word\_census.py}).

\paragraph{Answer extraction and benchmark validity.} That evaluations move under prompt formatting, and rankings with them, is well documented \cite{alzahrani,li2026,sclar}; we reproduce it in a financial setting and do not report it. Nor is the extraction rule's freedom ours to claim: against human judgment on multiple-choice QA, traditional strategies underestimate models and LLM-based extractors err systematically \cite{molfese}; first-token scoring is shown not to match the same model's text answer \cite{wang}; and \textsc{olmes} exists because formatting, few-shot choice and normalisation are unreported degrees of freedom that move a score \cite{gu}. What none supplies is a statistic. They measure the spread extraction choices produce; we report the interval a score is pinned to before any variant runs and count its width as $M^*$. We then trace the exposure to mechanisms: the echo (\S\ref{sec:echo}), a shipped \texttt{DEFAULT} that scores unanswerable generations at the floor (\S\ref{sec:floor}), and a fingerprint in published numbers (\S\ref{sec:exact}).

The closest work argues the optimistic case. Prompt sensitivity is attributed to ``log-likelihood scoring and rigid answer matching,'' with an LLM judge cutting variance across templates \cite{hua}, the recovery of mis-scored items offered as mechanism rather than measured. On short-answer VQA an audit measures it: up to half of penalised items are semantically acceptable, yet \textbf{82.56\%} of confirmed false negatives survive the deterministic repair \cite{ye}; \S\ref{sec:recover} reaches the same shape of conclusion. Two audits ask a different question. A review of 445 benchmarks assesses construct validity across a benchmark's whole lifecycle \cite{bean}, and item-level defect detection flags invalid questions from response patterns \cite{truong}; both ask whether a benchmark is built soundly, where we ask whether one reported number is a property of the model at all -- arithmetically, it is mostly parse rate (\S\ref{sec:floor}). Within finance, a review of 164 finance-LLM papers enumerates five biases, none of them baseline omission or harness configuration \cite{kong}, and recent ICAIF work shows the prompting paradigm changes financial-sentiment outcomes \cite{vamvourellis}.

\section{Experimental Design}\label{sec:design}
\paragraph{Factorial design.} The factorial crosses 6 open-weight models (DeepSeek-R1-Distill-Qwen-1.5B and -7B, Qwen2.5-7B and Qwen2.5-7B-Instruct, Hermes-3-Llama-3.1-8B, EuroLLM-9B-Instruct) with 5 FLARE tasks, 2 prompt arms and 4 budgets $b \in \{16, 32, 64, 128\}$, yielding 204 cells. The arms are \textbf{raw} (prompt passed verbatim, what the pinned harness does) and \textbf{chat} (the same prompt wrapped in the model's chat template, the convention of newer harnesses). Generation is greedy, so a cell is deterministic and the only variance is the configuration. Three populations recur, and no cell belongs to more than one pooled statistic. (i) \emph{The factorial}: the 204 cells above. (ii) \emph{The diagnostic} (\S\ref{sec:rule}): 46 cells at $b{=}128$ -- the factorial's 28 raw sweep-task cells with Qwen2.5-14B-Instruct added as a seventh model, a type census of 17,164 scored items, plus a fresh 18-cell chat-arm run of the six factorial models on the three ungated tasks; neither 14B model appears in the chat arm. (iii) \emph{The \S\ref{sec:floor} legs}: Qwen2.5-14B-Instruct and \textsc{phi}-4, excluded from every pooled factorial statistic; \textsc{phi}-4 appears only in the family leg.

\paragraph{Task selection.} We take every FLARE task that is (i) classification with a closed answer vocabulary, so a degenerate floor is well defined, and (ii) small enough to run at all four budgets across six models within our compute, plus Headlines, which fails (ii) but exercises the other extraction path: German credit ($n{=}200$), Australian credit ($n{=}139$), Financial PhraseBank ($n{=}970$), CIKM stock movement ($n{=}1{,}143$) and Headlines ($n{=}20{,}547$). Criterion (i) excludes generative tasks, about which we make no claim. Every task runs on its full test split; splits are label-sorted, so a head-slice's floor would not be the split's. Headlines runs at one budget ($b{=}32$), giving $4 \times 6 \times 2 \times 4 = 192$ sweep cells plus 12 Headlines cells. Floors come from the gold distribution of each full split, per-stratum where the dataset declares strata (\S\ref{sec:exact}) and pooled majority otherwise. Every cell number we report is an accuracy, whereas FLARE reports these tasks in F1-family metrics, so no cell of ours is comparable to a published score; \S\ref{sec:exact}, the single point of contact, compares in the published AvgF1. The factorial also crosses base models with chat templates and instruction-tuned models with raw prompts; dismissing such cells as pathological would presuppose the convention whose absence is the finding, and the harness applies the raw path to instruction-tuned models by default.

\paragraph{Provenance of the reported accuracies.} Our runs score all four sweep tasks with \texttt{Classification.process\_results}, transcribed from source, which is the shipped rule on PhraseBank alone, so Table~\ref{tab:anchored} reports the shipped value beside ours; neither is a reimplementation of the other. Throughout Table~\ref{tab:anchored}: coverage, \texttt{acc.} and the Wilson verdict are ours; \texttt{ship.} is what the class each task actually instantiates returns on the same generations, which is why the two agree only on PhraseBank; $M$ is over the five admissible rules of \S\ref{sec:rule}; the floor is the majority class; and the echo rate uses the full-split denominator, whereas \S\ref{sec:echo} pools over parsed generations. Coverage is counted from the retained per-item dumps every other quantity uses, not the sweep's aggregate \texttt{missing\_rate}. \S\ref{sec:rule} reports instead what the five admissible rules return; matcher-invariance forces them to one value wherever $M = 0$, our scoring returns it on 9 of the 10 such cells it covers, and the tenth, one item of 970, is diagnosed there (\texttt{paper/source\_convention.py} fails the build if a second cell deviates).

\paragraph{Statistical procedure.} Cell accuracies are proportions over the full split, with 95\% Wilson intervals where a single cell is quoted. Models in a cell are scored on identical items, so comparisons are by McNemar's test on discordant pairs rather than an unpaired $z$; the floor's per-item predictions are known exactly, so \S\ref{sec:floor} tests against it the same way and states the multiple-testing treatment there.

\section{The Unspecified Extraction Rule}\label{sec:rule}
\paragraph{Two shipped rules, one generation.} FLARE's substring path iterates the dataset card's \texttt{choices} list \emph{in list order} and returns the first entry occurring anywhere in the lowercased generation; EleutherAI's \textsc{lm-evaluation-harness}, which PIXIU forks, resolves the same situation by \emph{position}, its \texttt{MultiChoiceRegexFilter} taking the first match in an alternation of the choice strings \cite{biderman}. Both are defensible; neither is reported beside the number it produces. We transcribe both from source rather than reimplement them.

\paragraph{What the transcription check establishes.} The transcription is checked against our own run's \texttt{scored\_acc}, not the harness's: it agrees on \textbf{21} of 24 cells to within 0.01, internal consistency rather than agreement with the shipped harness. The three misses (worst \textbf{0.0647}, Qwen2.5-7B-Instruct on Australian credit) touch no PhraseBank cell and none is a parsing difference: both legs score with \texttt{flare.py}'s own \texttt{process\_results} (\texttt{pod/run\_full.py}), but as independent greedy runs batched at 96 and 32, so items sit behind different left padding and a handful decode differently---the rule-free gold-mention rate moves with the accuracy on all three. Neither implementation is at fault and no reported number changes (\texttt{paper/transcription\_diag.py}). The same holds for the single item separating Table~\ref{tab:anchored}'s 0.7629 from the admissible family's 0.7639 on the certified PhraseBank cell: our transcription returns 0.7639, the value all five rules return, so the shipped rule does not exit the certified family on real data; 0.7629 is the harness leg's own figure, its rule-free gold-mention rate differing by exactly the one item its accuracy does, and that leg retains no per-item record, so the item cannot be named. At $M = 0$ and coverage 1.000 every admissible rule is forced to one value, so two values cannot be two rules. \texttt{preflight/run\_their\_code.py} executes the harness's own \textsc{Headlines} scorer verbatim (\S\ref{sec:exact}); no classification equivalent exists without the full harness.

On identical generations the two disagree by up to \textbf{0.4216} accuracy, and not through the inadmissible path. On that cell and on all four whose PhraseBank score crosses the floor between the rules, the installed filter's bare-letter fallback resolves \textbf{zero} items, so it returns the same accuracy to four decimals whether that stage is enabled or not, and all \textbf{2,135} disagreements there arise between rules each reading a choice string the text contains.

That extraction choices move scores is known \cite{alzahrani,biderman}; those are \emph{measurements}: run the variants, report the spread. The interval a score is pinned to before any variant runs is inherited, and the next two paragraphs say from whom. Nor is this prompt-side option-order bias: the text is fixed and only its scoring changes.

\paragraph{What the rule leaves undetermined.} Leaving it unspecified leaves the reported number \emph{partially identified}. For a cell with generations $g_1 \ldots g_n$ and choice set $C$, fix the \emph{occurrence predicate} once ($c$ occurs in $g$ iff $c$ is a substring of $g$ after lowercasing, which is what both harnesses implement) and write the \textbf{candidate set} $S_i = \{c \in C : c \text{ occurs in } g_i\}$. A rule is \textbf{admissible} if (a) it returns an element of $S_i$ or nothing and (b) it returns nothing only when $S_i = \emptyset$; neither is free: the always-abstain rule satisfies (a) alone and scores 0 where every admissible rule scores 0.7639 (Qwen2.5-7B on PhraseBank, \S\ref{sec:floor}). Let the \textbf{ambiguity rate} $M$ be the fraction of items with $|S_i| \geq 2$, call item $i$ \emph{contested} when its gold label $y_i$ also lies in $S_i$, and let $M^*$ be the contested rate. Two consequences follow, stated without proof because neither is the contribution: \emph{matcher-invariance}, that any two admissible rules $A$ and $B$ satisfy $\bigl|\mathrm{acc}_A - \mathrm{acc}_B\bigr| \leq M$; and the \emph{reachable interval}, that the accuracies an admissible rule can attain form an interval $[\,\mathrm{lo}, \mathrm{hi}\,]$ of width $M^*$ with both endpoints attained.

\paragraph{What is inherited.} Sharpness is inherited, not established here: for a $0/1$ outcome the interval is the sharp identification region under coarsening \cite{artstein,manski1989,manski2003,molinari}, carried into LLM evaluation by Rao and Callison-Burch, who pin a rubric judge's full-set accuracy to an interval as wide as its uncovered fraction \cite{rao}. What differs is the coarsening: theirs is \emph{missingness}, the item abstained on and nothing recoverable; ours is \emph{multiplicity}, the harness reading one label when it could have read another. Multiplicity supplies a candidate set, which a missingness bound has no analogue of, and the set is what makes $M^*$ available: an ambiguous item whose candidates miss the gold is wrong under every admissible rule, so $M^* < M$ wherever one occurs, and $M^*$ is counted over contested items rather than optimised over rules; the instantiation and its measured magnitude are the contribution.

Condition (a) excludes rules that ship today. Run the \emph{installed} \textsc{lm-evaluation-harness} filter over our generations: on the \textbf{15,350} that name a choice its choice-string fallback decides, and our transcription reproduces it on \textbf{every one}. The other 1,814 name none, and there the filter splits. On \textbf{673} it still returns a label, through a bare-letter fallback keyed to the task's own choice letters, labels the text never names, so (a) fails and matcher-invariance is silent about them; the remaining \textbf{1,141} it declines to label at all. Rightly so: German credit's feature codes are \texttt{A11}, \texttt{A34}, \texttt{A65}, so \texttt{X1: A11} reads as answer \texttt{A}, a generation echoing the input features scored as if it had answered. An LLM-judge extractor \cite{hua} is excluded for the same reason.

So is a rule the suite itself ships: \texttt{StockMovement} resolves an empty candidate set through its synonym table on \textbf{459} of our 8,892 items and returns its \texttt{DEFAULT} on a further \textbf{819} -- the text names no choice, a label is returned regardless, (a) fails as for the bare-letter fallback. The invariance results are silent about it too; \S\ref{sec:floor} prices it.

Permuting \texttt{choices} is one rule under two orders; the two statistics nearly coincide (mean gap 0.0056 over the 28 raw-arm cells), so the label-free screen loses little. $M^*$ is what an unexamined rule could cost; the two shipped rules already realise it on PhraseBank.

\begin{figure}[t]
  \centering
  \includegraphics[width=240.25pt]{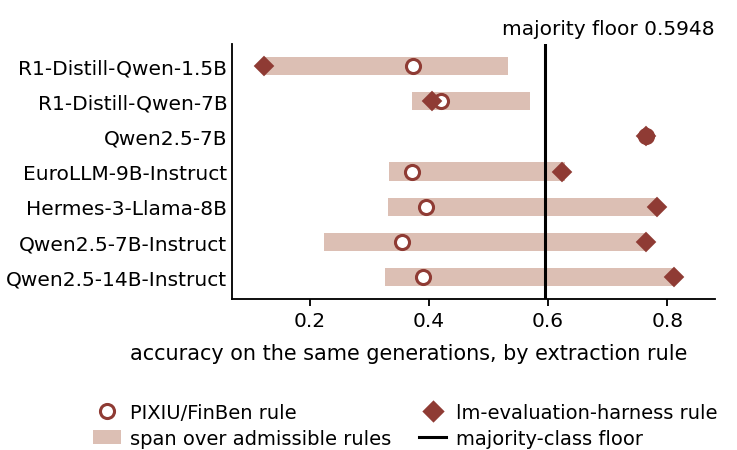}
  \caption{The same generations, scored by the two \emph{shipped} rules. Open circle: PIXIU/FinBen. Filled diamond: \textsc{lm-evaluation-harness}. Bar: span over the five admissible rules; values in Table~\ref{tab:phrasebank} (Appendix~\ref{app:tab}). Four of seven models sit below the majority-class floor under one rule and above it under the other.}
  \label{fig:rules}
\end{figure}

\paragraph{Validation.} Scoring five admissible rules over the \textbf{46 cells} (28 raw, 18 chat), matcher-invariance holds on every one and the bound is attained, so it cannot be improved without further assumptions. \textbf{Ten cells} sit at $M^* = 0$ (5 raw, 5 chat), and there every admissible rule returns \emph{identical} predictions. We call such a cell \emph{certified}, and the word means exactly this and no more: invariant over the admissible family at the configuration measured. Rules outside the family -- both fallbacks that ship -- are not covered, which is why \S\ref{sec:limits} leads with the case where that matters.

The certified count cannot stand alone. $M$ counts ambiguity among generations that name a label, so a cell naming none carries $M = 0$ because there is nothing for two rules to disagree about. \textbf{Four} of the ten are of that kind, all raw-arm at coverage below \textbf{0.05} (three parse nothing, the fourth four of 200), agreed accuracies \textbf{0.0000}--\textbf{0.0150}; coverage is computed in the same pass as $M$ and agrees with Table~\ref{tab:anchored}'s on all five cells. Only \textbf{one} of the five raw certified cells, Qwen2.5-7B on PhraseBank, is certified at full coverage. \emph{A certificate earned by silence is the same failure this paper diagnoses everywhere else in the pipeline}: a number determined by the extraction path rather than by the model. Coverage is the one quantity that separates the two, which is why \textsc{floorbench} prints coverage on the same line as the verdict and why rule 2 of \S\ref{sec:spec} is not optional beside rule 3. We therefore quote the certified count with its coverage split throughout.

\paragraph{Realised exposure and verdict reversal.} On PhraseBank the two \emph{shipped} rules reverse the reading of four of seven models, all on $n{=}970$ with gaps far larger than the Wilson width, on the same weights, generations and answer key (Figure~\ref{fig:rules}). Whether those four beat guessing is decided by which harness a reader runs, and neither answer is a mistake. PhraseBank is our only task where the exposure crosses a floor; permuting \texttt{choices} moves it by a median 0.4103 on PhraseBank against 0.0360 on the binary tasks, the binary figure separate because those prompts enumerate options in declared order, confounding metadata order with prompt copying. We claim only the floor crossings, and those are not noise.

\section{Question-Echo Scoring}\label{sec:echo}
FLARE prompts state the answer options inside the question, as in ``\ldots{} will \texttt{rise} or \texttt{fall} on 2017-11-02'', and the \texttt{Classification} path searches the generation for each label as a substring. A generation that merely \emph{restates the question} therefore contains a label string, parses successfully, and is scored, though the model has not answered: \texttt{To determine whether the closing price of GILD will rise or fall on 2017-11-02 based\ldots} is a verbatim stock-movement generation graded as a prediction.

We flag a scored generation as a question echo when the extracted label sits inside a disjunctive restatement of the options ($\pm$14 characters, tolerating the quotes the credit prompts use). Auditing the instrument -- strip every restatement, ask which label survives -- bounds its false positives, not its false negatives. In \textbf{1,055 of 4,077} flagged generations (25.9\%) none does, so the model never named an option on its own; in the other 3,022 it committed to a label and on \textbf{2,168 of those} (71.7\%) the harness graded a different one, discarding a real answer to grade a fragment of the prompt. Grading the model's own answer would flip 1,365 items to correct and 515 to wrong, net 850 against the model; \textbf{288} are wrong either way.

Echo rates are large and heterogeneous, and because the stripping test bounds only false positives, every rate below is a lower bound. Pooled over six models, 58.6\% of parsed German credit and 37.2\% of PhraseBank generations are echoes, spanning \textbf{200 of 200} down to \textbf{0 of 200} on German credit alone (Table~\ref{tab:anchored}'s echo column). The rate is not a stable model property: Hermes-3, at 0 of 200 there, echoes on \textbf{511 of 970} PhraseBank generations (52.7\%) while parsing every item on all four tasks, so its floor-level scores are a model answering and being wrong, and five of the six model rows range over at least \textbf{0.41} across the four tasks. The sixth, Qwen2.5-7B, is flat at zero -- none of 970 parsed on PhraseBank, \textbf{0.0061} of 1,143 on stock movement, its credit cells nearly empty -- and is the one model this mechanism does not reach. Echoing belongs to the (model, task) pair, so the rate cannot be corrected for once and forgotten.

Because \texttt{choices} is a fixed list, the extracted label is constant: all 1,338 stock-movement echoes return \texttt{rise}, and the other three tasks are 100\% constant as well. Echo scoring is a constant predictor laundered through an answer parser. Which constant is \texttt{choices[0]} -- fixed by the dataset card, not the data -- so whether it coincides with the majority class is an accident of wording. On stock movement and German credit it coincides, so echoes score $0.6001$ ($n{=}$\textbf{1,338}) and $0.6584$ (\textbf{401}), at their floors, invisible; on Australian credit and PhraseBank \texttt{choices[0]} is \texttt{good} and \texttt{positive}, neither the majority, and echoes score $0.4407$ (\textbf{177}) and $0.1948$ (\textbf{2,161}), far below. The four denominators pool model--item pairs over six models; no test is run on them. Echo does not reliably inflate a score toward the floor; it pins the score to a field the model never sees, landing above or below by accident.

Echo and \S\ref{sec:rule} are one phenomenon seen twice: a generation restating the question names \emph{every} option, so echo drives $M$ up and \texttt{choices} order converts that ambiguity into a score. Hence the extremes: DS-R1-1.5B at $M = 1.0000$ on German credit, Qwen2.5-7B certified at $M = 0$.

\section{Floor Census}\label{sec:floor}
This section asks how many cells beat a degenerate predictor at all, and the answer turns on a distinction the accuracy column cannot draw. Under \texttt{Classification}, which scores unparsed items wrong, reported accuracy factorises exactly as coverage $\times P(\text{correct} \mid \text{parsed})$ (the \texttt{StockMovement} override never emits \texttt{missing} and does not factorise this way), and the second factor is the substantive one. Over the 165 cells with coverage above 0.01 its unweighted per-cell mean is $0.5295$, on average \textbf{0.0710 below the task's majority-class floor}; item-weighted over the 90,296 parsed items it is $0.5473$, \textbf{0.0431 below the floor}. A cell can sit under its floor for either reason, and the two are priced differently. Charging an unparsed generation is a \emph{defensible convention}, so low coverage is not invalid, only \emph{unreported}. Invalidity is claimed for two things alone: items scored \emph{as though answered} when nothing was asserted (\S\ref{sec:echo}), and the absence of any floor.

\paragraph{Definition of clearing the floor.} A cell clears the floor iff the lower Wilson bound on its accuracy exceeds the floor, sits below iff the upper bound is under it, and ties otherwise. Reporting ties explicitly is the conservative direction for our own claim: on stock movement the best cell is 0.5792 (Hermes-3, chat arm, all four budgets) against a floor of 0.5809, which on $n = 1{,}143$ is a tie, not a loss.

Of the 192 sweep cells \textbf{15 clear their floor, 24 tie, and 153 sit below}; every one of the 15 is on Financial PhraseBank, and on the three binary tasks ($6 \times 8 \times 3 = 144$ cells) \textbf{not one clears}. They are not an artifact of testing 192 cells uncorrected, since at $\alpha = 0.05/192$ \textbf{all 15 survive} while \textbf{9} below-verdicts widen into ties. The Wilson verdict only ever says a cell fails to clear; the constant's per-item predictions are known exactly, so the comparison can be made paired, and by McNemar over the 24 cells of the anchored configuration \textbf{19} are significantly \emph{worse} than the constant and \textbf{1} beats it: Qwen2.5-7B on PhraseBank, 264 discordant pairs its way against 100. The marginals match Table~\ref{tab:anchored}'s 19 bel / 4 tie / 1 clr, with two cells exchanging between tie and below across the verdicts. The paired tests are uncorrected; at $\alpha = 0.05/24$ \textbf{15} of the 19 survive, \textbf{4} widen into ties, and the beating cell survives. On German and Australian credit the predictor it tests against is \texttt{StockMovement}'s \texttt{DEFAULT}, so the test and the shipped rule bear on the same object. Adding Headlines (\S\ref{sec:exact}, 12 further cells, none clearing), 4 of the 5 tasks admit no configuration under which any tested model is distinguishable from a degenerate predictor.

\begin{table}[t]
  \centering
  \scriptsize
  \setlength{\tabcolsep}{3pt}
  \begin{tabular}{@{}llrrrrrlrr@{}}
    \toprule
    model & task & $n$ & floor & cov. & acc. & ship. & v. & $M$ & echo\\
    \midrule
    DS-R1-1.5B & austr. & 139 & 0.5683 & 0.993 & 0.4317 & 0.4317 & bel & 0.9928 & 0.993\\
    DS-R1-7B$^{\ddagger}$ & austr. & 139 & 0.5683 & 0.511 & 0.2374 & 0.4532 & bel & 0.2374 & 0.237\\
    EuroLLM-9B$^{\dagger}$ & austr. & 139 & 0.5683 & 0.000 & 0.0000 & 0.4317 & bel & 0.0000 & 0.000\\
    Hermes-3-8B & austr. & 139 & 0.5683 & 1.000 & 0.4101 & 0.4029 & bel & 0.0144 & 0.007\\
    Qwen2.5-7B$^{\dagger}$ & austr. & 139 & 0.5683 & 0.000 & 0.0000 & 0.4317 & bel & 0.0000 & 0.000\\
    Qwen2.5-7B-Ins & austr. & 139 & 0.5683 & 0.719 & 0.3237 & 0.5324 & bel & 0.1007 & 0.036\\
    \addlinespace
    DS-R1-1.5B & fpb & 970 & 0.5948 & 0.998 & 0.3722 & 0.3722 & bel & 0.7546 & 0.469\\
    DS-R1-7B & fpb & 970 & 0.5948 & 0.986 & 0.4175 & 0.4175 & bel & 0.3196 & 0.247\\
    EuroLLM-9B & fpb & 970 & 0.5948 & 1.000 & 0.3711 & 0.3711 & bel & 0.5082 & 0.277\\
    Hermes-3-8B & fpb & 970 & 0.5948 & 1.000 & 0.3928 & 0.3928 & bel & 0.6165 & 0.527\\
    Qwen2.5-7B$^{\dagger}$ & fpb & 970 & 0.5948 & 1.000 & 0.7629 & 0.7629 & clr & 0.0000 & 0.000\\
    Qwen2.5-7B-Ins & fpb & 970 & 0.5948 & 0.998 & 0.3598 & 0.3598 & bel & 0.8144 & 0.707\\
    \addlinespace
    DS-R1-1.5B & germ. & 200 & 0.6600 & 1.000 & 0.6600 & 0.6600 & tie & 1.0000 & 1.000\\
    DS-R1-7B$^{\dagger}$ & germ. & 200 & 0.6600 & 0.000 & 0.0000 & 0.6600 & bel & 0.0000 & 0.000\\
    EuroLLM-9B & germ. & 200 & 0.6600 & 1.000 & 0.6600 & 0.6600 & tie & 0.9950 & 0.995\\
    Hermes-3-8B & germ. & 200 & 0.6600 & 1.000 & 0.3450 & 0.3500 & bel & 0.0050 & 0.000\\
    Qwen2.5-7B$^{\dagger}$ & germ. & 200 & 0.6600 & 0.020 & 0.0150 & 0.6750 & bel & 0.0000 & 0.000\\
    Qwen2.5-7B-Ins & germ. & 200 & 0.6600 & 0.400 & 0.2050 & 0.4850 & bel & 0.1050 & 0.010\\
    \addlinespace
    DS-R1-1.5B & sm & 1,143 & 0.5809 & 0.990 & 0.5039 & 0.5048 & bel & 0.4112 & 0.347\\
    DS-R1-7B & sm & 1,143 & 0.5809 & 0.968 & 0.5626 & 0.5748 & tie & 0.4602 & 0.414\\
    EuroLLM-9B & sm & 1,143 & 0.5809 & 0.876 & 0.4742 & 0.5416 & bel & 0.0586 & 0.026\\
    Hermes-3-8B & sm & 1,143 & 0.5809 & 1.000 & 0.5407 & 0.5442 & bel & 0.0700 & 0.007\\
    Qwen2.5-7B & sm & 1,143 & 0.5809 & 0.990 & 0.5783 & 0.5844 & tie & 0.0079 & 0.006\\
    Qwen2.5-7B-Ins & sm & 1,143 & 0.5809 & 0.849 & 0.4873 & 0.5687 & bel & 0.3858 & 0.370\\
    \bottomrule
  \end{tabular}
  \caption{The 24 anchored-configuration cells (raw arm, $b{=}128$); column conventions and sources in \S\ref{sec:design}; echo rates use the full-split denominator. $^{\dagger}$ the five certified cells ($M = 0$). $^{\ddagger}$ \texttt{acc.} and $M$ coincide at four decimals by coincidence (33/139 each). \textsc{Headlines} resolves by byte equality, so $M$'s occurrence predicate does not apply; its 12 cells are in \S\ref{sec:exact}.}
  \label{tab:anchored}
\end{table}

Table~\ref{tab:anchored} gives the per-cell view at the configuration we recommend anchoring to (raw arm, $b{=}128$; the \emph{anchored configuration} hereafter), every figure in it an accuracy. \textbf{1} of the 24 clears its floor: Qwen2.5-7B on PhraseBank at \textbf{0.7629}, which is also the one cell the diagnostic certifies at full coverage ($M = 0$); the single-item gap to the admissible family's 0.7639 is diagnosed in \S\ref{sec:rule}.

\begin{figure*}[t]
  \centering
  \includegraphics[width=\textwidth]{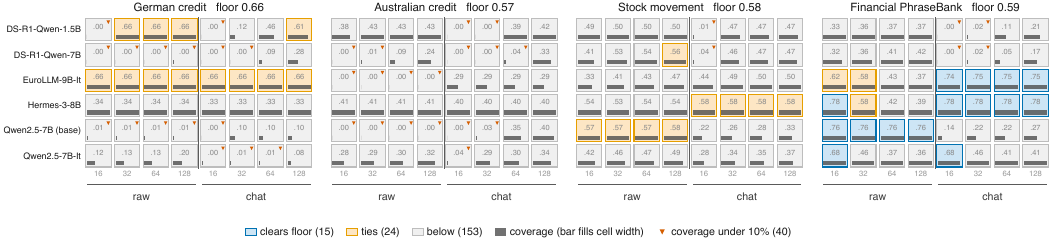}
  \caption{The 192 sweep cells by Wilson verdict against the task floor; bars give coverage, triangles mark coverage under 10\%. \textsc{Headlines} uses the exact-match path and is excluded.}
  \label{fig:census}
\end{figure*}

Figure~\ref{fig:census} supports two readings the aggregate counts alone cannot. First, survivors are \emph{model}-shaped, not configuration-shaped: every configuration column holds a non-below cell on every task that has one, while non-below cells cluster in two model rows on German credit and vanish on Australian. Second, against our own reading, \textbf{103 of the 153 below-floor cells sit under 90\% coverage}, 40 under 10\%: most of the census is models never producing a parseable answer, a fact about the pipeline rather than about financial competence.

\paragraph{What the shipped rule does with the same generations.} That DeepSeek cell parses \emph{nothing}: all 200 generations name no choice, so under \texttt{Classification} it scores \textbf{0.0000} at zero coverage. Under the rule \texttt{German} actually ships it scores \textbf{0.6600} at coverage \textbf{1.000} on the identical generations, the synonym table resolving one item and the \texttt{DEFAULT} -- \texttt{good}, the majority class -- supplying the other \textbf{199}, so the cell lands on the floor exactly. So does Qwen2.5-7B, from \textbf{0.0150} to \textbf{0.6750} on \textbf{187} defaulted items. A model emitting nothing parseable is not merely indistinguishable from one answering at the base rate; the benchmark's own code assigns it the base rate. Nor does this need the default to be the majority: on Australian credit it is the \emph{minority}, and EuroLLM-9B and Qwen2.5-7B each go from \textbf{0.0000} to \textbf{0.4317} with all 139 items defaulted. Across the three tasks the shipped rule moves a cell by up to \textbf{0.6600} and never reports a missing item: the \texttt{missing} flag tests \texttt{result} $=$ \texttt{missing}, and \texttt{result} is always a choice or the default, so the flag can never fire.

\begin{figure}[t]
  \centering
  \includegraphics{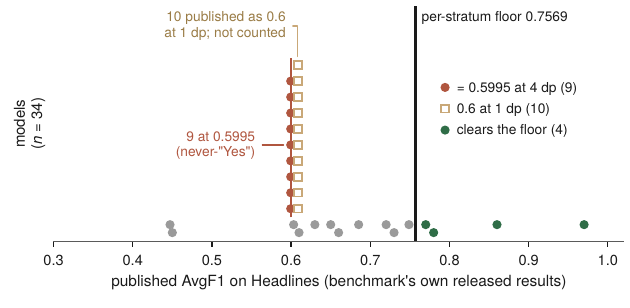}
  \caption{The benchmark's own released \textsc{Headlines} scores against the two floors its scoring code implies. Every value is an AvgF1, so the comparison never crosses metrics; the ten entries published at one decimal are drawn as open squares, offset for legibility.}
  \label{fig:headlines}
\end{figure}

\paragraph{Three causes separated on one task.} With the floor held fixed on German credit at the anchored configuration, three models occupy three different positions that one number cannot tell apart. Hermes-3 \textbf{answers and is beaten}, at coverage 1.00 with \textbf{zero} echoes and 0.3450 against the 0.6600 floor, an ordinary capability result no parser manufactured. Qwen2.5-14B-Instruct scores \emph{higher}, 0.4750, on \textbf{61.5\%} echoes. EuroLLM-9B-Instruct scores \textbf{0.6600}, exactly the floor, on \textbf{99.5\%} echoes, and ranks first. The accuracy order reverses the answering order. A floor-level score therefore cannot distinguish a model that failed from a score that was manufactured, and this corpus contains both. Nor is the mechanism confined to one model family: \textsc{phi}-4, off the Qwen lineage four of our six share, echoes on \textbf{95.6\%} of parsed stock-movement generations (\texttt{paper/family\_check.py}).

That EuroLLM-9B-Instruct is top-ranked on German credit under all eight configurations, at exactly 0.6600 every time, is why rank-invariance claims need a degeneracy guard: here the invariance is a floor-sitter's.

\section{The Exact-Match Path}\label{sec:exact}
If the invalidity were an artifact of the substring parser, \texttt{Headlines} should come out clean: byte equality leaves the matcher no freedom, so $M$ is identically zero. It does not, on our largest experiment, 246,564 generations.

Headlines is nine tasks, not one. Its \texttt{label\_type} column partitions the items into 9 Yes/No question types of 2,283 each with per-stratum majority rates from 0.5891 to 0.9930 (Table~\ref{tab:strata}, Appendix~\ref{app:tab}); the correct degenerate baseline is the per-stratum majority. Stratum assignment needs no test-label oracle, each item's question type being named verbatim in its prompt; the within-stratum majority comes from the gold distribution, as every floor here does (\S\ref{sec:design}). Always predicting ``No'' scores 0.6806 accuracy and 0.5995 AvgF1 (the metric FinBen reports: \texttt{Headlines.label\_avg}, the unweighted mean over the nine strata of \texttt{sklearn}'s support-weighted F1, with prediction 0 iff the output strips to \texttt{Yes}) against the per-stratum majority's 0.8285 and 0.7569, so measuring against the pooled constant credits $+0.1479$ accuracy of headroom no model earned. All 12 of our cells run the full $n{=}20{,}547$ split at coverage 1.000 and every one is below 0.8285 on its Wilson interval, \textbf{7 of the 12 scoring exactly 0.6806}, the pooled constant to four decimals, including cells with up to 19,264 distinct generations; the five not pinned there span 0.6427 to 0.7208.

\paragraph{The same signature in the published leaderboard.} Our runs could be dismissed as an audit of a commit current tooling has moved past; the published numbers cannot. Among published FLARE results this task is the one our audit reaches directly; the title's claim rests here, extending to the rest of the suite only through \S\ref{sec:background}'s census. We pulled this task's released per-model results from the benchmark's repository (the Hugging Face dataset \texttt{TheFinAI/results}, \texttt{demo-leaderboard/} at branch \texttt{main}; what is fixed is the cached copy we commit, \texttt{results/leaderboard\_headlines\_authoritative.csv}: 34 models, machine-readable, no transcription) and compared them against the floors its scoring code implies (Figure~\ref{fig:headlines}). Only four clear the per-stratum floor of 0.7569, which we compute from that code rather than take from anything the benchmark releases: FinMA-7B (0.97), GPT-4 (0.86), Gemini (0.78) and ChatGPT (0.77). Three are frontier API models we could not run; the fourth, FinMA-7B, is PIXIU's own release, instruction-tuned on the financial data built alongside this benchmark \cite{pixiu}, topping the released list ahead of GPT-4. Under the scoring-path reading this is the expected outcome -- a model tuned on the benchmark's own answer format is likeliest to emit the exact string the matcher demands -- but the capability reading is equally available: FinMA is fine-tuned on in-domain financial data and may simply be better at the task. We did not run FinMA-7B and cannot separate the two readings, which is the point: the column that would say so does not exist.

The sharper test is the exact value. The scorer returns 0.5995 when no generation is ever byte-equal to \texttt{Yes}, and \textbf{nine of the 34 are published at four decimals and equal it exactly}. The other route is narrow. Reproducing 0.5995 at four decimals without never-\texttt{Yes} requires a byte-exact-\texttt{Yes} rate in a window only 0.04\% of items wide around 42.0\% (\texttt{paper/yes\_rate\_band.py} derives both), whereas never-\texttt{Yes} is a single point every degenerate model hits by construction. Ten more are published as \texttt{0.6} at one decimal, a band non-degenerate patterns also occupy; we do not count them.

The roster cuts both ways: this is not an independent sample. \textbf{Four} of the nine we have no other evidence about (\textsc{starcoder2-7b}, \textsc{CodeLlama-7b}, \textsc{CodeLlama-7b-Instruct}, \textsc{DeepSeek-V3}); the other \textbf{five} (both R1 distills, \textsc{EuroLLM-9B-Instruct}, \textsc{Hermes-3-Llama-3.1-8B}, \textsc{Qwen2.5-7B-Instruct}) we run ourselves, and the check agrees: all five score exactly 0.6806 on our raw-arm rerun at a byte-exact-\texttt{Yes} rate of exactly \textbf{0}, while the sixth factorial model, Qwen2.5-7B, is off the constant on the released list (\textbf{0.7488}) and in our rerun (0.7138, byte-exact-\texttt{Yes} \textbf{0.3238}). The overlap costs independence and buys a mechanism check the other 25 entries cannot supply; what survives is a claim about numbers in the literature, not that nine unrelated groups reproduced the fingerprint.

\paragraph{The mechanism.} This path resolves to \texttt{Yes} only on byte-exact equality, so a model never emitting it has all 20,547 items scored \texttt{No}, giving exactly the base rate 0.6806. The data confirm this without exception: all 7 pinned cells have a byte-exact-\texttt{Yes} rate of \textbf{exactly 0} and all 5 unpinned cells a strictly positive one (0.027--0.515). Output diversity is thus no evidence against degenerate scoring on either path.

On an item whose gold is \texttt{Yes}, a pinned cell emits \texttt{Yes\textbackslash n\allowbreak </think>\textbackslash n\textbackslash nYes} verbatim, newlines shown as \texttt{\textbackslash n}: the model answers correctly, twice. The scorer strips the whitespace, compares the remainder to \texttt{Yes}, finds it unequal, and records \texttt{No}, the verdict it would record for a model that said nothing. This is the counterpart of the stock-movement echo of \S\ref{sec:echo}: there the parser reads an answer out of a question, here it reads no answer out of one that was given. Both yield a number that is a property of the extraction rule reported against no floor, so the failure is not specific to the substring parser.

\section{Recoverability of Unparsed Items}\label{sec:recover}
A natural objection holds that the fix is a better answer extractor. Part of the residue lies provably beyond every \emph{admissible} rule: of the \textbf{1,296} generations \texttt{Classification} marks \texttt{missing} across the six factorial models and four sweep tasks at $b{=}128$, \textbf{1,296 (100.0\%)} contain no label string anywhere, a census rather than a sample, so condition (a) leaves an admissible rule nothing to return. The shipped harness returns something on all but 18 of them: \textbf{1,278} fall on the three \texttt{StockMovement} tasks, where a synonym or the \texttt{DEFAULT} supplies a label and \texttt{missing} is never set (\S\ref{sec:floor}). That is a second rule outside the admissible family, not a recovery. The negative is universal over that family, not over rules that ship: on \textbf{649} lm-eval's bare-letter fallback returns a letter the generation never names -- the failure of \S\ref{sec:rule}, not an escape from it. On the \texttt{Headlines} path the repair is unambiguous -- truncate at the first newline, a one-line change the harness supports -- and rerunning the raw arm at $b{=}32$ splits its five pinned cells 2 to 3. Two recover genuine signal, reaching \textbf{0.7570} and \textbf{0.7061} against the pooled 0.6806 constant, and neither clears the per-stratum floor of \textbf{0.8285}; the other three \emph{lose} up to \textbf{0.2046}, swapping one constant for a worse one. Neither repair serves both paths, and we make no claim about repairs we did not test; the remedy we specify is reporting.

\section{The Reporting Specification}\label{sec:spec}
Four rules follow, ordered by implementation effort; none requires new generations. (1) State the prompt arm and generation budget, neither of which the benchmark specifies and both of which move the number. (2) Report coverage beside every accuracy; the harness computes and discards it. (3) Report the ambiguity rate $M$ and, with the answer key in hand, the reachable interval $[\,\mathrm{lo}, \mathrm{hi}\,]$. (4) Report the per-task floor, per-stratum where the dataset declares strata. Rule 2 separates failing from never answering; rule 3, whether the number survives a change of rule; rule 4, whether it beats guessing. On Qwen2.5-14B-Instruct's PhraseBank cell the three that are missing read \texttt{coverage 1.0000}, \texttt{M* 0.6340} and \texttt{floor 0.5948}, so the floor lies inside the reachable interval and the score cannot say whether it beats guessing. Where $M = 0$ the same call prints \texttt{certified} beside coverage, deliberately: a cell that parses nothing is certified too, and only the two fields together say which kind of certificate a reader holds (\S\ref{sec:rule}). The same line carries the card-order shipped rule's score, a function of what \textsc{floorbench} is already handed, plus any caller-supplied score under a rule it cannot reconstruct; where either moves a certified number the line names it and gives the size of the move, closing the gap \S\ref{sec:limits} demonstrates.

We release the specification at both points where it applies. \texttt{floorbench.py} computes all four offline from generations already produced, which makes the numbers recoverable for finished runs; \texttt{floorbench.patch} makes a run emit them itself, in \textbf{77 added lines across three files} that remove nothing, apply clean to both pinned commits, and change no existing metric. Agreement between the two implementations is checked rather than asserted, over \textbf{4,000} synthetic generations, to $10^{-12}$ on all five emitted quantities (\texttt{coverage}, $M$, $M^*$, \texttt{reach\_lo}, \texttt{reach\_hi}).

\section{Limitations}\label{sec:limits}
The diagnostic's scope is a restriction, not a formality. It covers rules returning a choice that occurs in the generation, so an LLM judge and lm-eval's bare-letter fallback both fall outside it, and the exclusion bites precisely where the certificate is proudest. DeepSeek-R1-Distill-Qwen-7B on German credit is certified at $M^* = 0$ and every admissible rule scores it \textbf{0.0000}; run the installed \textsc{lm-evaluation-harness} filter over the same generations and it returns \textbf{0.6550}, its bare-letter stage firing on \textbf{199} of the 200 items and reading the dataset's own feature codes as an answer. That gap of \textbf{0.6550} is larger than the two-rule gap this paper leads with (\S\ref{sec:rule}), on a cell where the rules the diagnostic covers agree to the last item, and it is not isolated: three of the five raw-arm certified cells move by more than 0.01 under the shipped filter (Qwen2.5-7B: 0.0150$\rightarrow$\textbf{0.6450} on German, 0.0000$\rightarrow$\textbf{0.4317} on Australian credit), all cells that parse almost nothing. Nothing here is wrong, since none of those resolutions is admissible and \S\ref{sec:rule} says so; what the example shows is that ``certified'' cannot be read as ``this number is safe'' without reading coverage in the same breath.

The open-weight restriction matters: of the 34 released entries, the four clearing the floor are three frontier API models and FinMA, while our evidence that a score can be manufactured is open-weight, 1.5B--14B. Mechanism analyses cover four tasks at one budget: only the raw-arm $b{=}128$ dumps are retained per item, while generations alone survive the chat and Headlines reruns and the family leg -- enough for the diagnostic's 46 cells and the census's 204. Budgets stop at $b{=}128$, small for reasoning-distilled models, but the dumps test the truncation reading directly: \textbf{118} of the 1,296 missing generations (9.1\%) end inside an unterminated reasoning span, while \textbf{962} (74.2\%) come from models emitting no reasoning span at all, so truncation reaches at most a tenth of the residue and the below-floor census is not a budget artifact (\texttt{paper/truncation\_census.py}). Financial PhraseBank is absent from the chat-arm rerun alone: that rerun builds fresh prompts and the \texttt{TheFinAI/flare-fpb} dataset is gated to our access token, so a third party cannot rerun that arm either; everything else we report for it is computed offline from retained generations. The harness premise, that the raw arm is what the benchmark implements, is verified against one pinned commit, and the published results of \S\ref{sec:exact} came from that pipeline, so its behaviour governs an existing body of reported numbers whatever a rerun would give. We audit measurement, not the datasets.

\section{Conclusion}\label{sec:concl}
A benchmark score is a measurement only to the extent it does not depend on which answer-extraction rule the harness happens to ship. The ambiguity rate $M$ bounds that dependence in one pass over generations the harness already produced, and $M\,{=}0$ certifies the number as invariant to the extraction rule at the configuration measured. Three numbers recoverable from artifacts the harness already produces resolve the question: coverage, $M$, and the floor.

\paragraph{Reproducibility.} Every number is regenerated from committed results by released scripts and checked by a provenance gate; generation is greedy, so cells are deterministic. That gate, \texttt{paper/numbertrace.py}, re-executes the recomputation scripts, fails on a nonzero exit rather than silently losing their values, matches every asserted value against their pooled output, and binds claims value matching cannot separate to named artifact fields. A companion self-test plants known errors and requires the gate to catch every one, so it licenses manuscript--artifact agreement, not artifact correctness.

\clearpage   
\appendix
\section{Tabulated Values}\label{app:tab}

\begin{table}[H]
  \caption{PhraseBank accuracies under the two shipped rules on identical generations (the values Figure~\ref{fig:rules} plots), with the admissible-family span; rows in the figure's order.}
  \label{tab:phrasebank}
  \centering
  \scriptsize
  \begin{tabular*}{200pt}{@{}l@{\extracolsep{\fill}}rrrr@{}}
    \toprule
    model & PIXIU rule & lm-eval rule & lo & hi\\
    \midrule
    DS-R1-1.5B & 0.3742 & 0.1227 & 0.1227 & 0.5330\\
    DS-R1-7B & 0.4196 & 0.4052 & 0.3722 & 0.5701\\
    Qwen2.5-7B & 0.7639 & 0.7639 & 0.7639 & 0.7639\\
    EuroLLM-9B & 0.3711 & 0.6227 & 0.3340 & 0.6289\\
    Hermes-3-8B & 0.3948 & 0.7835 & 0.3309 & 0.7835\\
    Qwen2.5-7B-Ins & 0.3546 & 0.7649 & 0.2237 & 0.7649\\
    Qwen2.5-14B-Ins & 0.3897 & 0.8113 & 0.3258 & 0.8113\\
    \bottomrule
  \end{tabular*}
\end{table}

\begin{table}[H]
  \caption{The nine \textsc{Headlines} strata: \texttt{label\_type}, $n$, and per-stratum majority rate. \texttt{label\_type} values are verbatim from the dataset, misspelling included.}
  \label{tab:strata}
  \centering
  \scriptsize
  \begin{tabular}{@{}lrr@{}}
    \toprule
    \texttt{label\_type} & $n$ & majority rate\\
    \midrule
    Price or Not & 2,283 & 0.8519\\
    Direction Up & 2,283 & 0.5891\\
    Direction Down & 2,283 & 0.6299\\
    Direction Constant & 2,283 & 0.9549\\
    PastPrice & 2,283 & 0.8134\\
    FuturePrice & 2,283 & 0.9615\\
    PastNews & 2,283 & 0.8590\\
    FutureNews & 2,283 & 0.9930\\
    Asset Comparision & 2,283 & 0.8038\\
    \bottomrule
  \end{tabular}
\end{table}

\end{document}